\documentclass[11pt,thmsa,fleqn]{article}
\usepackage{amsthm, amsmath, amssymb, epsfig, multicol,enumerate,float}

\usepackage{appendix}
\usepackage{setspace}
\usepackage{rotating}
\usepackage{ccaption}
\usepackage{latexsym}
\usepackage{appendix}
\usepackage{subfigure}
\usepackage{longtable}
\usepackage{natbib}
\usepackage{anysize}
\usepackage{rotating}
\usepackage{url}
\usepackage{dcolumn,longtable}
\usepackage[none]{hyphenat}
\marginsize{3 cm}{3 cm}{2.5 cm}{2.5 cm}

\renewcommand{\thefootnote}{\fnsymbol{footnote}}

\numberwithin{equation}{section}

\usepackage[T1]{fontenc}

\begin{document}

\title{{\LARGE { Optimal Trading Execution with Nonlinear Market Impact:\\ An Alternative Solution Method }}}

\author{{\large Massimiliano Marzo, Daniele Ritelli and Paolo Zagaglia}\footnote{Marzo: Department of Economics, Universit\`{a} di Bologna; massimiliano.marzo@unibo.it. Ritelli: Department of Statistics, Universit\`{a} di Bologna; daniele.ritelli@unibo.it. Zagaglia: Department of Economics, Universit\`{a} di Bologna; paolo.zagaglia@unibo.it.\newline {\bf Remarks:} This is a work in progress. We welcome any comments, suggestions and remarks. The research contained herein has been carried out without financial support of any kind from the private sector.}}

\date{{\small This version: \today}}
\singlespacing

\maketitle

\thispagestyle{empty}

\setcounter{footnote}{0}
\renewcommand{\thefootnote}{\arabic{footnote}}

\begin{abstract}
We consider the optimal trade execution strategies for a large portfolio of single stocks proposed by \citet{Al}. This framework accounts for a nonlinear impact of trades on average market prices. The results of \citet{Al} are based on the assumption that no shares of assets per unit of time are trade at the beginning of the period. We propose a general solution method that accomodates the case of a positive stock of assets in the initial period. Our findings are twofold. First of all, we show that the problem admits a solution with no trading in the opening period only if additional parametric restrictions are imposed. Second, with positive asset holdings in the initial period, the optimal execution time depends on trading activity at the beginning of the planning period.  \\ \\
Keywords: optimal execution, market impact, ordinary differential equations.\\
JEL classification: G11, G12.\\
\end{abstract}

\newpage
\setcounter{page}{1}

\onehalfspacing

\section{Introduction}

The execution of large trades in financial markets requires the balance between risks and costs. The main risk concerns the lack of availability of a counterparty, which can lead to a delay in the execution of a transaction. In order to guarantee a fast trade execution, a trader may incur additional costs. As clarified by \citet{HS_1}, a trader faces a choice between a `passive' and an `actice' execution strategy. 

Given this background, the available models of optimal execution assume that the trading activity of individual investors has an impact on the average price prevailing in the market. The transaction costs are characterized by parametric forms that replicate stylized facts documented in the market microstructure literature \citep[e.g. see][]{KSS}. 

Almgren and Chriss (1999, 2000) and Konishi and Makimoto (2001) provide examples of optimal strategies for the execution problem in the stock market. Their models assume that the transaction cost per share is a linear function of the number of shares of assets traded. The only source of uncertianty consists in the volatility of the stock price. 

\citet{Al} suggests that the linearity assumption is largely at odds with reality. First, the average liquidity premium on stocks tends to be either a convex or a concave function of the traded size. This depends on the counterparty's perception about the reason for the trade, namely on whether it is driven by liquidity or information needs \citep[see][]{HS}. Moreover, the liquidity premium is related to the risk of finding a counterparty. In other words, the lower the probability of finding a counterparty in the market, the higher the liquidity premium. 

In this paper, we review the optimal transaction strategy proposed by \citet{Al}. We show that the solution method used by \citet{Al} is ill-posed. The reason is that it is based on the assumption that no shares per unit of time are exchanged at the beginning of the period. We use an approach based on the Gauss hypergeometric function to solve for the case of positive initial trades. Our results differ strongly from those of \citet{Al}. First of all, the problem admits a solution with no trading in the opening period only if additional parametric restrictions are imposed. Second, with positive initial trading, the optimal execution time depends on trading activity in the initial period. 

This note is organized as follows. Section 2 provides a selected discussion of the literature on optimal trade execution. Section 3 outlines the structure of the problem.  Section 4 proposes a general solution method for positive initial values of the velocity. Section 5 concludes. Finally, in Appendix A, we discuss the general method for the solution of second order differential equations with a Gauss hypergeometric function.

\newpage

\section{A selected overview of the literature}

Rebalancing portfolios of assets requires executing trades in the marketplace. With the advent of algorithmic trading and access availability to many alternative trading venues, investors have dedicated increasing resources to the scheduling of trades. The practical setup of the problem is rather intuitive. An investor has a target number of, say shares that it intends to sell or buy within a given time frame. The decision problem requires to compute how many shares to place or demand in the market at each point in time within the trade horizon. The aim of the investor is to minimize the execution costs. These are typically measured as the difference between the price obtained from the market and a benchmark price for the transaction. 

There are multiple relevant dimensions to the execution problem. Several contributions have showed that the liquidity premium is time-varying. The reason is that it is determined by the availability of traders willing to act as counterparties, namely traders willing to buy or sell a given quantity of an asset at a desired price. However, as the presence of traders willing to `take the other side' of a trade is uncertain, any trading is characterized by execution risk. 

Another relevant aspect is related to the fact that market illiquidity generates transaction costs.  This typically takes the form of a large spread between bid and ask prices \citep{HS}. Therefore, as noticed by \citet{WB}, the minization of transaction costs is a key aspect of the portfolio optimization problem. As documented in various studies including \citet{Cha}, \citet{HLM} and \citet{KSS}, large trades do impact market prices and, thus affecting the bid-ask spreads. 

Asset price volatility is a source for execution risk. The reason is that it affects the probability of finding a suitable counterparty. Hence, it affects the sucessfulness of a trading strategy execution. The recent literature has focused on the specific aspect of volatility, namely the  increased uncertainty in execution price incurred by rapid execution of large share blocks.   
In fact, \citet{HSS} show that liquidity fluctuates due to intrinsic variations in market activity independently of trade size. 
 
Based on the considerations outlined earlier, what are the properties of an `optimal' execution strategy? What defines an `optimal execution price'? \citet{BL} argue that `best' execution can be thought of as a dynamic strategy that minimizes `liquidation' costs. They show that dynamic programming techniques can be used fruitfully to characterize these strategies.

Almgren and Chriss (2000) compute optimal trajectories for trading prices that are obtained by balancing market impact costs. The optimal profiles provide a motivation for low execution speed.\footnote{The literature has proposed two main alternative bechmarks. These consist of average prices that materialize within the trading horizon, and are characterized as a time-weighted and volume-weighted average price.} This results arises from the balance between the need to reduce the expected value of execution, and the need to minimize the adverse effects of market volatility. While the first factor provides a reason for slower execution, the second factor lays the ground for rapid execution. That would, in fact, reduce execution risk in the form of the variance of execution cost with respect to the benchmark price. In short, early execution reduces execution risk, whereas a delayed execution is more geared towards minimizing execution costs. Evidently the degree of investor risk aversion determines how early within the trading horizon the execution starts.  The shape of the schedule instead depends on the form of the assumed market impact model. 



\citet{KM} makes the assumption that the market impact of trading is a linear function of trade size. In this case, the optimality frontier representing the combinations of minimized costs and market volatility has an analytical solution. Value-at-risk utility funcations are then used to select the first-best solution. This choice of utility function provides a natural testing ground for the concept of liquidity-adjusted VaR, which explicity considers the best trade-off between volatility risk and liquidation cost. \citet{Al} generalizes the results of \citet{KM} to the case of nonlinear functions for the market impact of trades. In this framework, the assumption is that the market impact cost per share follows a power law function of the trading rate.

\section{The optimal execution problem}

We follow the general framework of Almgren and Chriss (2000). At time $t=0$, an investor holds $X$ shares of an asset. The problem is to sell these shares by by time $t=T$. We should stress that this is the statement of a general framework. In fact, the initial size $X$ can either be positive or negative. In the first case, the investor needs to schedule a selling program. In the second case, the investor looks at a buying program. In this paper, for simplicity, we focus on the case $X>0$. 

The execution problem consists in minizing the market impact of trades subject to both initial and terminal conditions. In mathematical terms, the model proposed by \citet{Al} delivers the following optimization problem:
\begin{equation}\label{P}
\left\{\begin{split}
&\min_{x(t)}\int_0^TF(x,\dot{x})\,{\rm d}t\\
&x(0)=X,\quad x(T)=0
\end{split}\right.
\end{equation}
where $F(x,y)$ is the market impact function of trades:
\begin{equation}
F(x,y)=-\gamma xy+\eta (-y)^{k+1}+\lambda\sigma^2x^2,\quad \gamma,\,\eta,\,\lambda,\,k>0.
\end{equation}
The problem is to determine the optimal function $x(e)$ so as to minimize a chosen cost functional. Using Beltrami identity, \citep[see][section 5, page 31]{KS}:
\begin{equation}
F(x,\dot{x})-\dot{x}\frac{\partial F}{\partial y}(x,\dot{x})={\rm constant}
\end{equation}
evaluating the constant of integration at the end time $T$ we are lead to the differential equation
\begin{equation}\label{E}
\lambda\sigma^2-k\eta(-\dot{x})^{k+1}=-k\eta(-\dot{x}(T))^{k+1} 
\end{equation}

\cite{Al} proposes a solution based only on the \lq\lq elementary'' case $v_0=0$. In this note, we show how to compute the constant $v_0^{k+1}:=(-\dot{x}(T))^{k+1} $ using the initial condition $x(0)=X$.

\subsection{The case $k=1$}

The case $k=1$ is straightforward since it gives rise to a linear ordinary differential equation, whose solution is better found starting from the classical Euler Lagrange equation:
\begin{equation}
\frac{{\rm d}}{{\rm d}t}\frac{\partial L}{\partial y}(x,\dot{x})=\frac{\partial L}{\partial x}(x,\dot{x})\iff 2 \lambda  \sigma ^2 x-2 \eta  \ddot{x}=0
\end{equation}
imposing the boundary conditions $x(0)=X,\,x(T)=0$ we find:
\begin{equation}\label{S1}
x(t)=X\frac{\sinh \left(\frac{\sqrt{\lambda }   }{\sqrt{\eta }}\,\sigma(T-t)\right)}{\sinh\left(\frac{ \sqrt{\lambda }}{\sqrt{\eta }}\,\sigma T\right)}
\end{equation}
It is worth noting that for $k=1$  \eqref{S1} gives a minimizer of \eqref{P} since in this case Legendre condition reads:
\begin{equation}
\frac{\partial^2F}{\partial y^2}(x,\dot{x})=2\eta>0
\end{equation} 
Turning back to equation \eqref{E} for general $k$, we write it solving with respect to $\dot{x}:$
\begin{equation}\label{E'}
\begin{cases}
\dot{x}=-\left(v_0^{k+1}+\dfrac{\lambda\sigma^2}{k\eta}x^2\right)^{\frac{1}{k+1}}\\
x(0)=X
\end{cases}
\end{equation}
In this general case observe that, since:
\begin{equation}
\frac{\partial^2F}{\partial y^2}(x,\dot{x})= \eta k(k+1)(-\dot{x})^{k-1}
\end{equation}
since solution to \eqref{E'} is decreasing we infer, for the Legendre condition, the minimality of the estremal $x(t).$

\section{A general solution method for the case of positive initial trades }

In this section,  we propose a solution method that holds when there are positive stock trades in the initial period, namely for $v_0>0$. This solution $x$ to \eqref{E'} is implicitely defined  by
\begin{equation}\label{E''}
\int_{x}^X\left(v_0^{k+1}+\dfrac{\lambda\sigma^2}{k\eta}z^2\right)^{-\frac{1}{k+1}}{\rm d}z=t.
\end{equation}
Integral in the right hand side of \eqref{E''} can be evaluated by means of the Gauss hypergeometric function $_2{\rm F}_1$ whose definition and basic properties are given in the appendix. After some changes of variables which allows to rewrite \eqref{E''} as:
\begin{equation}
\frac{1}{2v_0}\left\{X\int_{0}^{1}\frac{s^{-\frac12}}{\left(1+\frac{\lambda\sigma^2X^2}{k\eta v_0^{k+1}}s^2\right)^{\frac{1}{k+1}}}\,{\rm d}s-x\int_{0}^{1}\frac{s^{-\frac12}}{\left(1+\frac{\lambda\sigma^2x^2}{k\eta v_0^{k+1}}s^2\right)^{\frac{1}{k+1}}}\,{\rm d}s\right\}=t
\end{equation}
we can finally use the integral representation Theorem, see equation \eqref{R} in the appendix, for the hypergeometric function to obtain
\begin{equation}\label{Sk}
\frac{1}{v_0}\left\{X\,_2{\rm F}_1\left( \left. 
\begin{array}{c}
\frac12,\,\frac{1}{k+1} \\[1mm] 
\frac32
\end{array}
\right|- \frac{\lambda\sigma^2X^2}{k\eta v_0^{k+1}}\right)-x\,_2{\rm F}_1\left( \left. 
\begin{array}{c}
\frac12,\,\frac{1}{k+1} \\[1mm] 
\frac32
\end{array}
\right|- \frac{\lambda\sigma^2x^2}{k\eta v_0^{k+1}}\right)\right\}=t
\end{equation}
To obtain $x(t)$ from equation \eqref{Sk} observe that the function 
\begin{equation}
x\mapsto x\,_2{\rm F}_1\left( \left. \begin{array}{c}
\frac12,\,\frac{1}{k+1} \\[1mm] 
\frac32
\end{array}
\right|-\frac{\lambda\sigma^2x^2}{k\eta v_0^{k+1}}\right)
\end{equation}
is strictly decreasing function for values of the independent variable $>0$ and so is possible to revert and obtain $x(t)$ from \eqref{Sk}, if needed, numerically or, better, using the Lagrange power series reversion when:
\begin{equation}
\left|\frac{\lambda\sigma^2x^2}{k\eta v_0^{k+1}}\right|<1
\end{equation}
But in \eqref{Sk} there is no determination of $v_0$ which is essential for the full solution of the problem. If we limit to assign some specific values for $v_0$ as in the \lq\lq easy case'' $v_0=0$ we lose control on the initial value $x(0)=X.$ 

The way to obtain the final velocity in order to fit with  the initial value $x(0)=X$ is explained below. We use the so called \lq\lq shooting method'' as presented for instance in \cite {Bu} section 7.3.1 pages 502-507. The starting point is the Euler Lagrange equation with initial values in $T$
\begin{equation}\label{EL}
\begin{cases}
\ddot{x}=\dfrac{2\lambda\sigma^2}{\eta k(k+1)}\,x(-\dot{x})^{1-k}\\
x(T)=0,\quad \dot{x}(T)=-v_0
\end{cases}
\end{equation}
To solve \eqref{EL} we use the change of variables $u:=x,\,y:=\dot{x}$ following \cite{mur} section 2.3 pages 160-161 and, since 
\begin{equation}
\ddot{x}=y\frac{{\rm d}y}{{\rm d}u}
\end{equation}
equation \eqref{EL} is transformed in
\begin{equation}\label{EL'}
\begin{cases}
\dfrac{{\rm d}y}{{\rm d}u}=-\dfrac{2\lambda\sigma^2}{\eta k(k+1)}\,(-y)^{-k}\,u\\
y(0)=-v_0
\end{cases}
\end{equation}
Since \eqref{EL'} is separable we can integrate it, so that, returning back to the original variables we find:
\begin{equation}\label{EL''}
\begin{cases}
\dot{x}=-\left(v_0^{k+1}+\dfrac{\lambda\sigma^2}{\eta k}\,x^2\right)^{\frac{1}{k+1}}\\
x(T)=0
\end{cases}
\end{equation}
Equation \eqref{EL''} is separable and the relevant integration needs again the hypergeometric integral. Discarding some tedious computations we find out that solution to \eqref{EL} is defined implicitely by:
\begin{equation}\label{ELs}
\frac{x}{v_0}\,_2{\rm F}_1\left( \left. 
\begin{array}{c}
\frac12,\,\frac{1}{k+1} \\[1mm] 
\frac32
\end{array}
\right|- \frac{\lambda\sigma^2x^2}{k\eta v_0^{k+1}}\right)=T-t
\end{equation}
In order to meet the second initial condition $x(0)=X$ we see that $v_0$ must satisfy:
\begin{equation}\label{v0}
\frac{X}{v_0}\,_2{\rm F}_1\left( \left. 
\begin{array}{c}
\frac12,\,\frac{1}{k+1} \\[1mm] 
\frac32
\end{array}
\right|- \frac{\lambda\sigma^2X^2}{k\eta v_0^{k+1}}\right)=T
\end{equation}
Note that, being assigned all parameters $\lambda,\,\sigma,\,X,\,\eta,\,k$ equation \eqref{v0} is an equation in the sole unknown $v_0$. Of course such equation has to be treated numerically: once the value of $v_0$ is detected it has te be inserted in \eqref{Sk} to obtain solution to \eqref{P}.

As an (easy) example take $\lambda=\sigma=X=\eta=T=1$ and $k=1/2.$ In the plot below we represent the left hand side of \eqref{v0} and the value of $T$ at the right hand side. Using Mathematica$^{\text{\textregistered}}$ we find numerically $v_0=0.671525.$

\begin{figure}[H]
\begin{center}
\scalebox{1}{\includegraphics{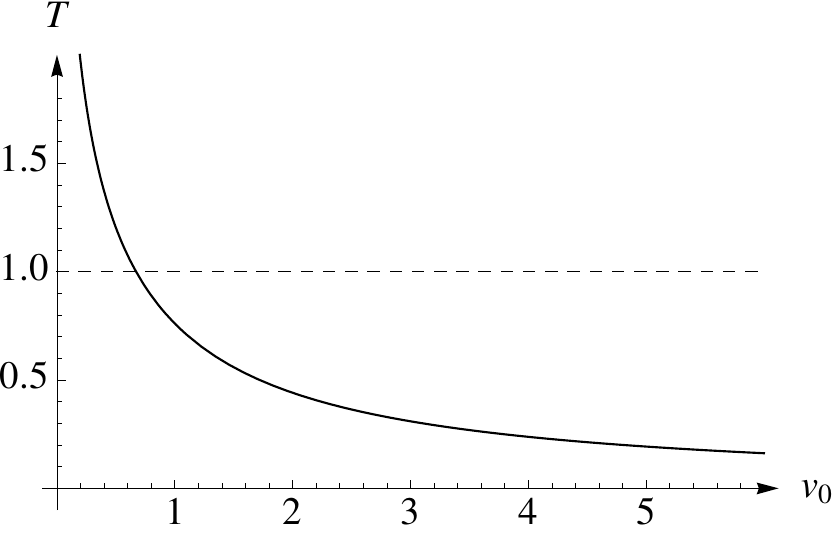}}
\caption{Plot of equation \eqref{v0}}
\end{center}
\end{figure}

Afterwards we put this value in \eqref{Sk} and we plot the relevant function $x(t)$ 
\begin{figure}[H]
\begin{center}
\scalebox{1}{\includegraphics[scale=0.95]{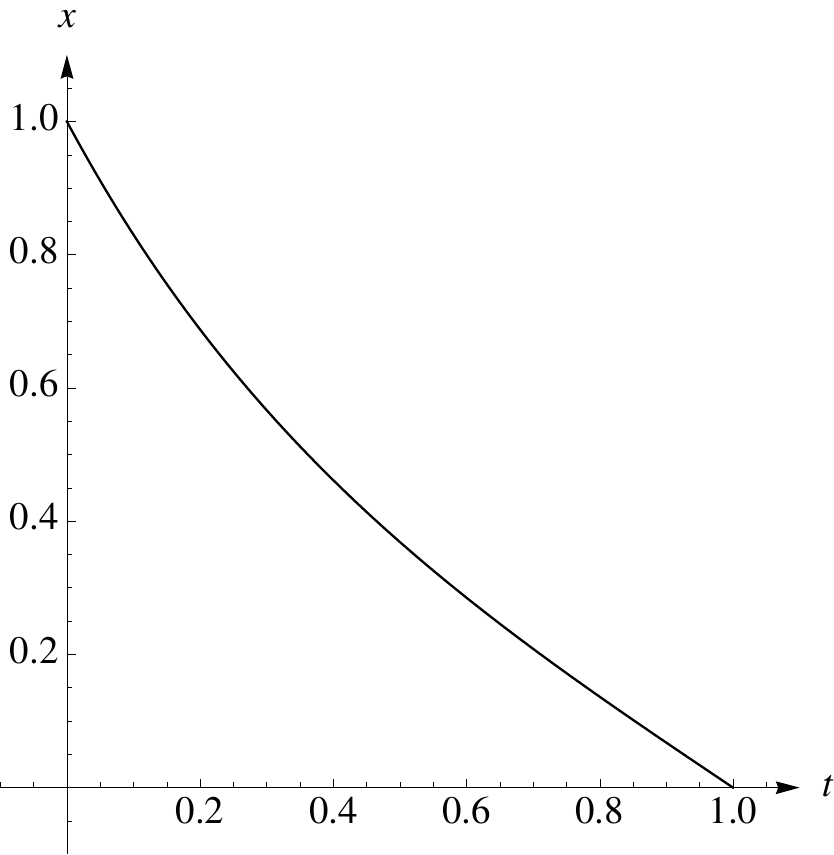}}
\caption{Plot of solution to \eqref{P}}
\end{center}
\end{figure}

We conclude with a graphic representation, with the same parameters for several values of $k$.
\begin{figure}[H]
\begin{center}
\scalebox{1}{\includegraphics[scale=0.95]{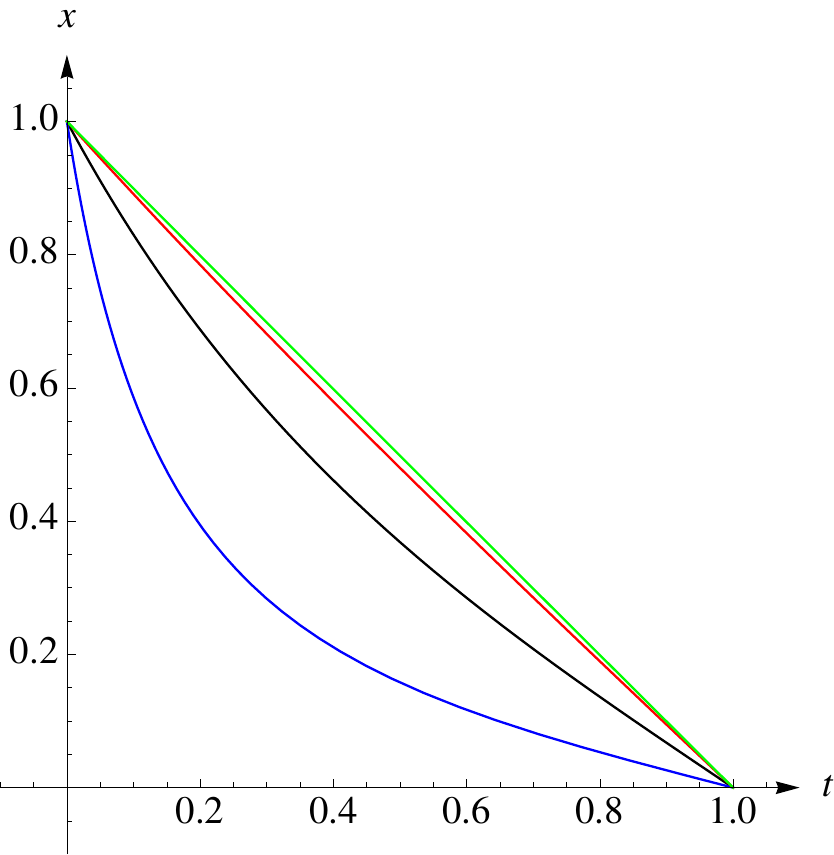}}
\caption{Solution to \eqref{P} $k=1/8$ (blue), $k=1/2$ (black), $k=2$ (red), $k=8$ (green)}
\end{center}
\end{figure}

\subsection{The Almgren zero-speed case}

When Almgren takes $v_0=0$ the initial value problem \eqref{EL} has to be considered with zero initial conditions $x(T)=\dot{x}(T)=0$. Observe that the quadrature formula arising from \eqref{EL''}, which is equivalent to \eqref{EL} reads in this case as:
\begin{equation}\label{zeros}
\int_0^{x}\frac{{\rm d}z}{\left(v_0^{k+1}+\frac{\lambda\sigma^2}{\eta k}z^2\right)^{\frac{1}{k+1}}}=T-t\implies\int_0^{x}\frac{{\rm d}z}{\left(\frac{\lambda\sigma^2}{\eta k}\right)^{\frac{1}{k+1}}z^{\frac{2}{k+1}}}=T-t
\end{equation}
but this means that the convergence condition $\frac{2}{k+1}<1$ has to be imposed. Moreover since we assume $k$ to be positive this means that the zero speed assumption is well posed only if $k>1$: for $k\leq1$ there are not solutions of the optimization problem \eqref{P} with zero speed. Moreover if we evaluate the integral at the left hand side of \eqref{zeros} for $k>1$ we find:
\begin{equation}
\frac{k+1}{k-1}\left(\frac{\lambda  \sigma ^2}{k \eta
   }\right)^{-\frac{1}{k+1}} x^{\frac{k-1}{k+1}}=T-t\implies x(t)=\left(\frac{(k-1) (T-t)}{k+1}\right)^{\frac{k+1}{k-1}}
   \left(\frac{\lambda  \sigma ^2}{k \eta }\right)^{\frac{1}{k-1}}
\end{equation}
so that at $t=0$ we have:
\begin{equation}
x(0)=\left(\frac{(k-1) T}{k+1}\right)^{\frac{k+1}{k-1}} \left(\frac{\lambda 
   \sigma ^2}{k \eta }\right)^{\frac{1}{k-1}}
\end{equation}
this means that we are not free to consider an arbitrary value of $x(0)$ having assigned the speed in $t=T$.

\subsection{The case $k=1$}

In the case $k=1$ we have provided two solutions of \eqref{P}: the first follows from the straightforward integration of the linear case, see equation \eqref{S1}, while the second stems from the hypergeometric implicit solution expressed by equation \eqref{Sk}. Of course the two solution to \eqref{P} are, as matter of fact, the same. This can be understood recalling the following property of $_2{\rm F}_1,$ \citep[see][entry 15.1.7, page 556]{AS}:
\begin{equation}
_2{\rm F}_1\left(\left.\frac12,\frac12\atop \frac32\right|-z^2\right)=\frac{\ln\left(z+\sqrt{1+z^2}\right)}{z}=\frac{{\rm arcsinh}\,z}{z}.
\end{equation}
In such a way, in this particular case we can use this identity to solve \eqref{ELs} with respect to $t$ obtaining:
\begin{equation}\label{sh1}
x(t)=\frac{v_0 \sinh \left((T-t) \sqrt{\frac{\lambda  \sigma ^2}{\eta 
   k}}\right)}{\sqrt{\frac{\lambda  \sigma ^2}{\eta  k}}}
\end{equation}
To compare this hypergeometric solution with \eqref{S1} we have to evaluate $v_0$ from \eqref{v0}:
\begin{equation}
v_0=\frac{\sqrt{\lambda } \sigma  X}{\sqrt{\eta } \sqrt{k}\sinh \left(\frac{\sqrt{\lambda } \sigma  T}{\sqrt{\eta } \sqrt{k}}\right)}
\end{equation}
and substitute in \eqref{sh1} getting:
\begin{equation}
x(t)=X\frac{\sinh \left(\frac{\sqrt{\lambda } \sigma  (T-t)}{\sqrt{\eta } \sqrt{k}}\right)}{\sinh \left(\frac{\sqrt{\lambda } \sigma  T}{\sqrt{\eta } \sqrt{k}}\right)}
\end{equation}
which is noting else but \eqref{S1}.

\newpage

\section{Conclusion}

Rebalancing large portfolios of stocks requires taking into account two peculiar issues. The first one is related to the market impact of trades, which generates transaction costs. The second issue arises from the risk of finding counterparties willing to trade at the desired price. Both empirical and theoretical considerations suggest that the market impact of trades is typically nonlinear. \citet{Al} proposes an optimal execution strategy that minimizes the tradeoff between volatility risk and transaction costs while taking into account this form of nonlinearity. 

In this paper, we review the optimal liquidation strategy of \citet{Al}. We show that the solution method used by \citet{Al} is ill-posed. The reason is that it is based on the assumption that no shares per unit of time are traded at the beginning of the period. We use an approach based on the Gaussian hypergeometric function to solve for the case of positive initial trades. Our results differ strongly from those of \citet{Al}. First of all, the problem admits a solution with no trading in the opening period only if additional parametric restrictions are imposed. Second, with positive initial trading, the optimal execution time depends on trading activity in the initial period.

\newpage

\begin{appendix}

\section{Solution of second order differential equations with a Gauss hypergeometric function}

The linear second order differential equation in the unknown $u=u(t)$
\begin{equation}\label{hy}
t(1-t)\,\ddot{u}+\left[c-\left(a+b+1\right)t\right]\,\dot{u}-ab\,u=0
\end{equation}
is known as Gauss hypergeometric equation. Parameters $a,\,b,\,c$ are not functions of the independent variable $t$ and can be in general complex number. Searching for a power series solution of \eqref{hy} it can be seen that
\begin{equation}
_2{\rm F}_1\left(\left.a,b\atop c\right|t\right):=\sum_{n=0}^\infty\frac{(a)_n(b)_n}{(c)_n}\,\frac{{t^n}}{{n!}}
\end{equation}
where we use the Pochhamer symbol $(x)_n,\,n\in\mathbb{N}$ defined as:
\begin{equation}
\begin{cases}
(x)_0:=1\\
(x)_n=x(x+1)(x+2)\cdots(x+n-1)
\end{cases}
\end{equation}
is the solution of \eqref{hy} such that $u(0)=1,\,\dot{u}(0)=ab/c.$ Power series defining $_2{\rm F}_1$ converges for $|t|<1$ and to continue the hypegeometric function $_2{\rm F}_1$ it is useful the integral representation ascribed to Leonhard Euler but really due to Adrien Marie Legendre\footnote{\textit{Exercices de calcul int\'{e}gral}, II, quatri\'{e}me part, section 2, Paris,  1811.}:
\begin{equation}\label{R}
{_{2}\mathrm{F}_{1}}\left( \left. 
\begin{array}{c}
a,\,b \\ 
c
\end{array}
\right| t\right) =\frac{\Gamma (c)}{\Gamma (c-a)\Gamma (a)}\int_{0}^{1}\frac{
s^{a-1}(1-s)^{c-a-1}}{(1-ts)^{b}}\mathrm{d}s, 
\end{equation}
where $\mathrm{Re}\,c>\mathrm{Re}\,a>0,\,|t|<1,$ and the Euler-Legendre integral (Gamma function) is defined for $x>0$ by: 
\begin{equation}
\Gamma (x)=\int_{0}^{\infty }e^{-u}u^{x-1}\mathrm{d}u.
\end{equation}
A proof of the integral representation theorem and a good presentation of the Gauss hypergeometric function can be found at \cite{Sea}, the integral representation theorem is treated at section 10.7, pages 184-185, formula (10.39). Integral representation theorem provides an extension to the region where the complex hypergeometric function is defined, namely for its analytical continuation, to the (almost) whole complex plane excluding the half-straight line $(1,\infty).$ This function was first introduced in dynamical economics in a generalization of the Solow Swan model due to \cite{R}, while \cite{Bou} use it in the Lucas-Uzawa model.

\end{appendix}



\begin{thebibliography}{9}

\bibitem[Abramowit and Stegun(1964)]{AS} Abramowitz, M. and I. Stegun (1964), {\it Handbook of Mathematical Functions}, Dover, New York.

\bibitem[Almgren(2003)]{Al} Almgren, R. F. (2003), ``Optimal execution with nonlinear impact functions and trading-enhanced risk'', {\it Applied Mathematical Finance}, 10, 1-18.

\bibitem[Almgren and Chriss(1999)]{Al_Ch_1} Almgren, R. and Chriss, N. (1999), ``Value under liquidation'', {\it Risk}, 12:(12), 61-3.

\bibitem[Almgren and Chriss(2000)]{Al_Ch_2} Almgren, R. and Chriss, N. (2000), ``Optimal execution of portfolio transactions'', {\it Journal of Risk}, 3:(2), 5-39.

\bibitem[Bertsimas and Lo(1998)]{BL} Bertsimas, D., and A. W. Lo (1998), ``Optimal control of liquidation costs'', {\it Journal of Financial Markets}, 1, 1-50.


\bibitem[Boucekkine and Ruiz-Tamarit(2008)]{Bou} Boucekkine, R. and J.R. Ruiz-Tamarit (2008), ``Special functions for the study of economic dynamics: The case of the Lucas-Uzawa model'', {\it Journal of Mathematical Economics}, {44}, 33-54.

\bibitem[Chakravarty(2001)]{Cha} Chakravarty, S. (2001), ``Steath-trading: which traders' trades move prices?, {\it Journal of Financial Economics}, 61, 289-307.


\bibitem[Hasbruock and Schwartz(1988)]{HS_1} Hasbrouck, J. and R. A. Schwartz (1988), ``Liquidity and execution costs in equity markets'', {\it Journal of Portfolio Management}, 14:(Spring), 10-16.

\bibitem[Hasbruock and Seppi(2001)]{HSS} Hasbrouck, J. and Seppi, D. J. (2001). ``Common factors in prices, order flows, and liquidity'', {\it Journal of Financial Economics}, 59, 383-411.

\bibitem[Holthausen, Leftwich and Mayers(1990)]{HLM} Holthausen, R. W., R. W. Leftwich and D. Mayers, D. (1990), ``Large-block transactions, the speed of response, and temporary and permanent stock-price effects'', {\it Journal of Financial Economics}, 26, 71-95.

\bibitem[Huang and Stoll(1997)]{HS} Huang, R. D. and H. R. Stoll (1997), ``The components of the bid-ask spread: a general approach'', {\it Review of Financial Studies}, 10:(4), 995-1034.

\bibitem[Jones and Lipson(1999)]{JL} Jones, C. M. and M. L. Lipson (1999), ``Execution costs of institutional equity orders'', {\it Journal of Financial Intermediation}, 8, 123-40.


\bibitem[Kamien and Scwartz(1991)]{KS} M.I. Kamien, N. L Schwartz (1991), {\it Dynamic Optimization. The Calculus of Variations and Optimal Control in Economics and Management}, Second edition, North Holland, Amsterdam.


\bibitem[Konishi and Makimoto(2001)]{KM} Konishi, H. and N. Makimoto (2001), ``Optimal slice of a block trade'', {\it Journal of Risk}, 3:(4).

\bibitem[Kraus and Stoll(1972)]{KSS} Kraus, A. and H. R. Stoll (1972), ``Price impacts of block trading on the New York Stock Exchange'', {\it  Journal of Finance}, 27, 569-88.


\bibitem[Mingari Scarpello and Ritelli(2003)]{R} Mingari Scarpello, G.  and D. Ritelli (2003), ``The Solow model improved through the logistic manpower growth law'', {\it Annali Universit\`{a} di Ferrara} - Sez VII - Sc. Mat., Vol. IL, 73-83.

\bibitem[Murphy(1960)]{mur} Murphy, G. M. (1960), {\it Ordinary differential equations and their solutions}, Van Nostrand Reinholds Company, New York.

\bibitem[Seaborn(1991)]{Sea} Seaborn, J. B. (1991), {\it Hypergeometric functions and their applications}, Springer-Verlag, New York.

\bibitem[Stoer and Burlish(1993)]{Bu} Stoer, J. and R. Burlish (1993), {\it Introduction to numerical analysis}, second edition, Springer-Verlag, New York.

\bibitem[Wagner and Banks(1992)]{WB} Wagner, W. H. and Banks, M. (1992) ``Increasing portfolio effectiveness via transaction cost management'', {\it Journal of Portfolio Management}, 19, 6-11.


\end{thebibliography}
\end{document}